\documentclass[12pt]{article}
\usepackage[]{graphicx}
\usepackage[]{color}
\makeatletter
\def\maxwidth{ %
	\ifdim\Gin@nat@width>\linewidth
	\linewidth
	\else
	\Gin@nat@width
	\fi
}
\makeatother

\definecolor{fgcolor}{rgb}{0.345, 0.345, 0.345}

\usepackage{framed}
\makeatletter
{\par\unskip\endMakeFramed%
	\at@end@of@kframe}
\makeatother

\definecolor{shadecolor}{rgb}{.97, .97, .97}
\definecolor{messagecolor}{rgb}{0, 0, 0}
\definecolor{warningcolor}{rgb}{1, 0, 1}
\definecolor{errorcolor}{rgb}{1, 0, 0}
\newenvironment{knitrout}{}{} 

\usepackage{alltt}

\usepackage[latin1]{inputenc}
\usepackage{amsfonts}
\usepackage{amssymb}
\usepackage{amsmath}
\usepackage{amsthm}
\usepackage{graphicx,psfrag,epsf}
\usepackage{enumerate}
\usepackage{bbm}
\usepackage{float}
\usepackage[authoryear,sort&compress]{natbib}
\newcommand{\blind}{0}

\IfFileExists{upquote.sty}{\usepackage{upquote}}{}
\begin{document}
	\def\spacingset#1{\renewcommand{\baselinestretch}%
		{#1}\small\normalsize} \spacingset{1}

	\large
	
	\newcommand{\al}{\mbox{$\alpha$}}
	\newcommand{\be}{\mbox{$\beta$}}
	\newcommand{\ep}{\mbox{$\epsilon$}}
	\newcommand{\gam}{\mbox{$\gamma$}}
	\newcommand{\sig}{\mbox{$\sigma$}}
	
	\DeclareRobustCommand{\FIN}{%
		\ifmmode 
		\else \leavevmode\unskip\penalty9999 \hbox{}\nobreak\hfill
		\fi
		$\bullet$ \vspace{5mm}}
	
	\newcommand{\calA}{\mbox{${\cal A}$}}
	\newcommand{\calB}{\mbox{${\cal B}$}}
	\newcommand{\calC}{\mbox{${\cal C}$}}
	
	\newcommand{\muas}{\mbox{$\mu$-a.s.}}
	\newcommand{\Nat}{\mbox{$\mathbb{N}$}}
	\newcommand{\Rea}{\mbox{$\mathbb{R}$}}
	\newcommand{\Prob}{\mbox{$\mathbf{P}$}}
	\newcommand{\ProbQ}{\mbox{$\mathbf{Q}$}}
	
	\newcommand{\nin}{\mbox{$n \in \mathbb{N}$}}
	\newcommand{\suc}{\mbox{$\{X_{n}\}$}}
	\newcommand{\sucP}{\mbox{$\mathbb{P}_{n}\}$}}
	
	\newcommand{\conv}{\rightarrow}
	\newcommand{\convn}{\rightarrow_{n\rightarrow \infty}}
	\newcommand{\convp}{\rightarrow_{\mbox{c.p.}}}
	\newcommand{\convs}{\rightarrow_{\mbox{a.s.}}}
	\newcommand{\convw}{\rightarrow_w}
	\newcommand{\convd}{\stackrel{\cal D}{\rightarrow}}
	\newcommand{\R}{\mathbb{R}}
	\newcommand{\Rn}{\mathbb{R}^n}
	\newcommand{\PR}{\mathbb{P}}
	\newcommand{\Rd}{{\displaystyle\mathbb{R}^2}}
	\newcommand{\Rb}{\mathbb{\overline{R}}}
	\newcommand{\Rbd}{{\displaystyle \mathbb{\overline{R}}^2}}
	\newcommand{\Rbn}{{\displaystyle \mathbb{\overline{R}}^n}}
	\newcommand{\I}{\mathbb{I}}
	\newcommand{\Sv}{\mathbb{S}}
	\newcommand{\Id}{{\displaystyle\mathbb{I}^2}}
	\newcommand{\In}{{\displaystyle\mathbb{I}^n}}
	\newcommand{\Z}{\mathbb{Z}}
	\newcommand{\N}{\mathbb{N}}
	\newcommand{\Hbb}{\mathbb{H}}
	\newcommand{\lp}{\left(}
	\newcommand{\rp}{\right)}
	\newcommand{\lc}{\left[}
	\newcommand{\rc}{\right]}
	\newcommand{\lb}{\left\{}
	\newcommand{\rb}{\right\}}
	\newcommand{\lf}{\left.}
	\newcommand{\ri}{\right.}
	\newcommand{\id}{\stackrel{d}{=}}
	\newcommand{\prob}[1]{\PR\lp#1\rp}
	\newcommand{\esp}[2]{\mathbf{E}_#1\lc#2\rc}
	\newcommand{\espe}[1]{\mathbf{E}\lc#1\rc}
	\newcommand{\cb}{C\!\!\!\!/\!\!\!/\,}
	\newcommand{\notprec}{\not\prec}
	\providecommand{\abs}[1]{\left|#1\right|}
	\newcommand{\casos}[4]{\lb\begin{array}{ll}#1,&#2\\#3,&#4\end{array}\ri}
	\newcommand{\caso}[2]{\lb\begin{array}{l}#1\\#2\end{array}\ri}
	\newcommand{\sign}{\textrm{sign}}
	\newcommand{\corr}[2]{\textrm{Corr}\lp#1,#2\rp}
	\newcommand{\cov}[2]{\textrm{Cov}\lp#1,#2\rp}
	\newcommand{\Cov}[1]{\textrm{Cov}\lp#1\rp}
	\newcommand{\var}[1]{\textrm{Var}\lp#1\rp}
	\newcommand{\lcm}{\text{lcm}}
	\setlength{\parindent}{0in}
	
	\newcommand{\lrp}[1]{\left(#1\right)}
	\newcommand{\lrc}[1]{\left[#1\right]}
	\newcommand{\lrb}[1]{\left\{#1\right\}}
	\newcommand{\E}[1]{\mathbf{E}\lc #1\rc}
	\newcommand{\V}[1]{\mathbb{V}\mathrm{ar}\lc #1\rc}
	\newcommand{\Es}[2]{\mathbf{E}_{#2}\lc #1\rc}
	\newcommand{\Vs}[2]{\mathbb{V}\mathrm{ar}_{#2}\lc #1\rc}
	\newcommand{\mse}[1]{\mathrm{MSE}\lrc{#1}}
	\newcommand{\mise}[1]{\mathrm{MISE}\lrc{#1}}
	\newcommand{\amise}[1]{\mathrm{AMISE}\lrc{#1}}
	\newcommand{\pf}[2]{\frac{\partial #1}{\partial #2}}
	\newcommand{\pftwo}[2]{\frac{\partial^2 #1}{\partial #2^2}}
	\newcommand{\pfmix}[3]{\frac{\partial^2 #1}{\partial #2\partial #3}}
	\newcommand{\norm}[1]{\left|\left| #1\right|\right|}
	\newcommand{\tr}[1]{\text{tr}\left[#1\right]}
	\newcommand{\inprod}[2]{\langle#1,#2\rangle}
	\newcommand{\vlinel}[1]{\multicolumn{1}{|c}{#1}}
	\newcommand{\vliner}[1]{\multicolumn{1}{c|}{#1}}
	
	\newcommand{\bb}[1]{\mathbb{#1}}
	\newcommand{\mcal}[1]{\mathcal{#1}}
	\newcommand{\mat}[1]{\mathbf{#1}}
	\newcommand{\ind}[1]{\mathbbm{1}_{\lrb{#1}}}

	\newtheorem {Prop}{Proposition} [section]
	\newtheorem {Lemm}[Prop] {Lemma}
	\newtheorem {Tabla}[Prop] {Table}
	\newtheorem {Theo}[Prop]{Theorem}
	\newtheorem {Coro}[Prop] {Corollary}
	\newtheorem {Nota}{Remark}[Prop]
	\newtheorem {Ejem}[Prop] {Example}
	\newtheorem {Defi}[Prop]{Definition}
	\newtheorem {Figu}[Prop]{Figure}
	

	\if0\blind
	{
		\title{\bf The DD$^G$-classifier in the functional setting}
		\author{J.A. Cuesta--Albertos\thanks{
				Research partially supported by the Spanish Ministerio de Ciencia y Tecnolog\'{\i}a, grants  
			 MTM2011-28657-C02-02 and MTM2014-56235-C2-2-P }\hspace{.2cm}\\
			Dept. of Mathematics, Statistics and Computation, Univ. of Cantabria, Spain\\
			M. Febrero--Bande\thanks{Research partially supported by the Spanish Ministerio de Ciencia e Innovaci\'{o}n, grants MTM2013-41383-P} and M. Oviedo de la Fuente$^\dagger$ \\
			Dept. of Statistics and Op. Res. Univ. of Santiago de Compostela, Spain} 
		\maketitle
	} \fi
	
	\if1\blind
	{
		\bigskip
		\bigskip
		\bigskip
		\begin{center}
			{\LARGE\bf The DD$^G$-classifier in the functional setting}
		\end{center}
		\medskip
	} \fi
	
	\bigskip

\begin{abstract}
The Maximum Depth classifier was the first attempt to use data depths instead of multivariate raw data in classification problems. Recently, the DD--classifier has fixed some serious limitations of this classifier but some issues still remain. This paper is devoted to extending the DD--classifier in the following ways: first, to be able to handle more than two groups; second, to apply regular classification methods (such as $k$NN, linear or quadratic classifiers, recursive partitioning,\ldots) to DD--plots, which, in particular, allows to obtain useful insights through the diagnostics of these methods; and third, to integrate various sources of information (data depths, multivariate functional data,\ldots) in the classification procedure in an unified way. An enhanced revision of several functional data depths is also proposed. A simulation study and applications to some real datasets are also provided.
\end{abstract}

\spacingset{1.45}

\section{Introduction}
In this paper we explore the possibilities of the depths in classification problems in multidimensional or functional spaces. Depths are, relatively simple, tools intended to  order  the points in a space depending on how deep they are with respect to a probability distribution, \Prob. 

In the one-dimensional case, it is easy to order points with respect to  \Prob , with the median being the innermost point and the extreme percentiles the outermost ones. Moreover, if $F_P$ denotes the distribution function of \Prob , then
\begin{equation} \label{Eq.Depth1}
D_P(x) = \min \{F_P(x),1-F_P(x)\}
\end{equation}
is an index which measures how deep $x \in \Rea$ is with respect to  \Prob . This index can also be applied to samples replacing $F_P$ by the empirical distribution function. Other possibilities for defining $D_P(x)$ are available (see, for instance, Subsection \ref{subs.FMD}), including those  in which  $D_P(x)$  decreases with the distance between $x$ and the mean of \Prob , which, in turn, is the deepest point. Most of them  are positive and bounded, and the bigger the index,   the deeper the point.

In the multidimensional case there exists no natural order; thus, ordering the points from the inner to the outer part of a distribution or sample is not so easy. To overcome this difficulty, several depths have been proposed using different approaches. A nice review of multivariate depths is \cite{Liu1999}. 

To the best of our knowledge, the first paper in which depths were used for classification was \cite{Liu1990}, where the MD-classifier (MD-classifier) was proposed: given two probability measures  (or classes, or groups) \Prob \ and \ProbQ, and a depth, $D$, we classify the point $x$ as produced by \Prob \ if $D_P(x) > D_Q(x)$. This procedure was fully developed in \cite{Ghosh2005}.

\begin{figure}
\centering
\begin{knitrout}\footnotesize
\definecolor{shadecolor}{rgb}{0.969, 0.969, 0.969}\color{fgcolor}

{\centering \includegraphics[width=.75\linewidth]{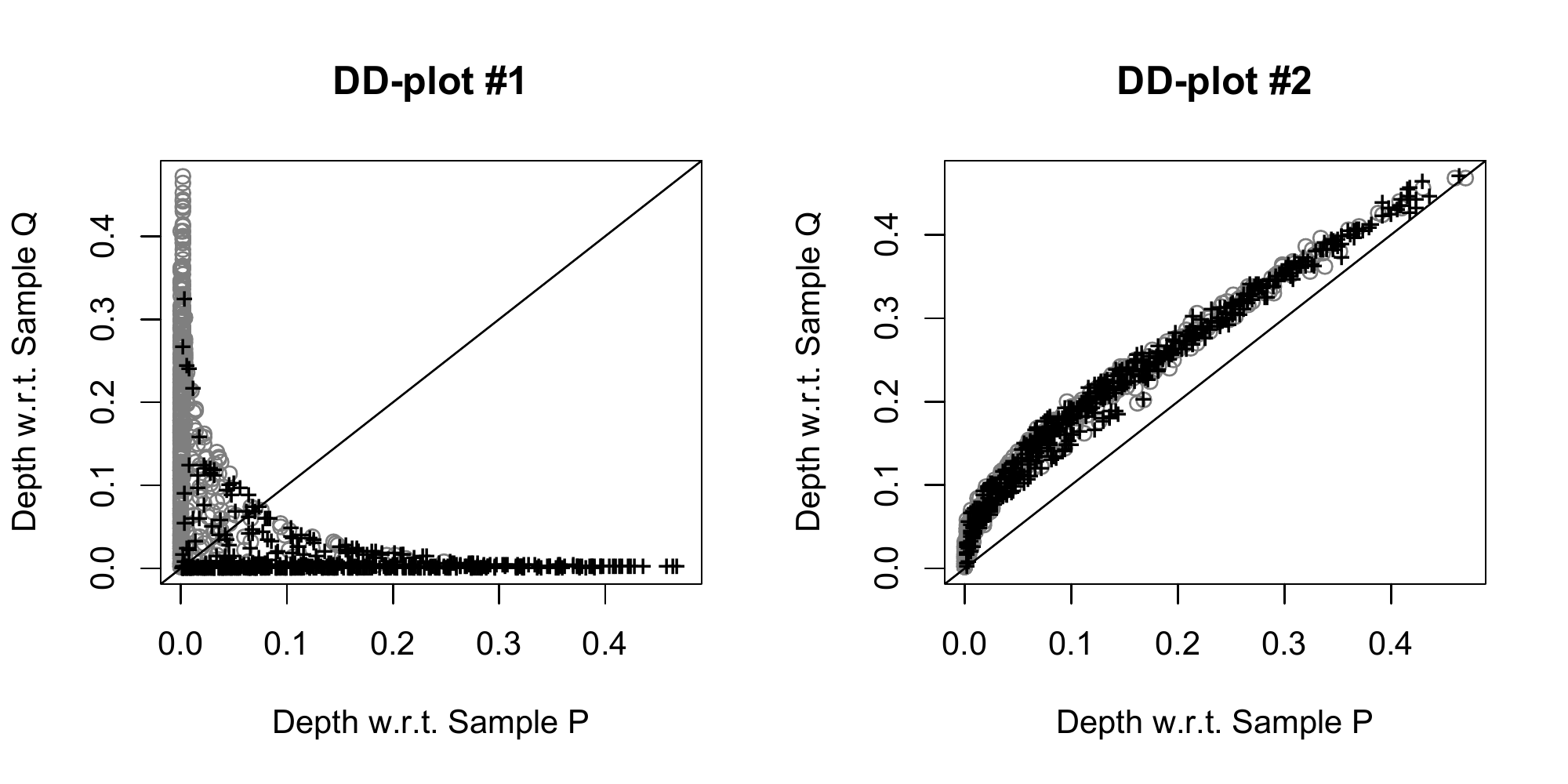} 

}

\end{knitrout}

\caption{DD--plots of two samples drawn from two-dimensional normal distributions. In both cases \Prob \ is a standard 2-dimensional distribution. \ProbQ \ differs from \Prob\ in the mean in the first DD--plot and in the covariance matrix in the second.}
\label{Fig1}
\end{figure}

\begin{figure}
\centering
\begin{knitrout}\footnotesize
\definecolor{shadecolor}{rgb}{0.969, 0.969, 0.969}\color{fgcolor}

{\centering \includegraphics[width=.75\linewidth]{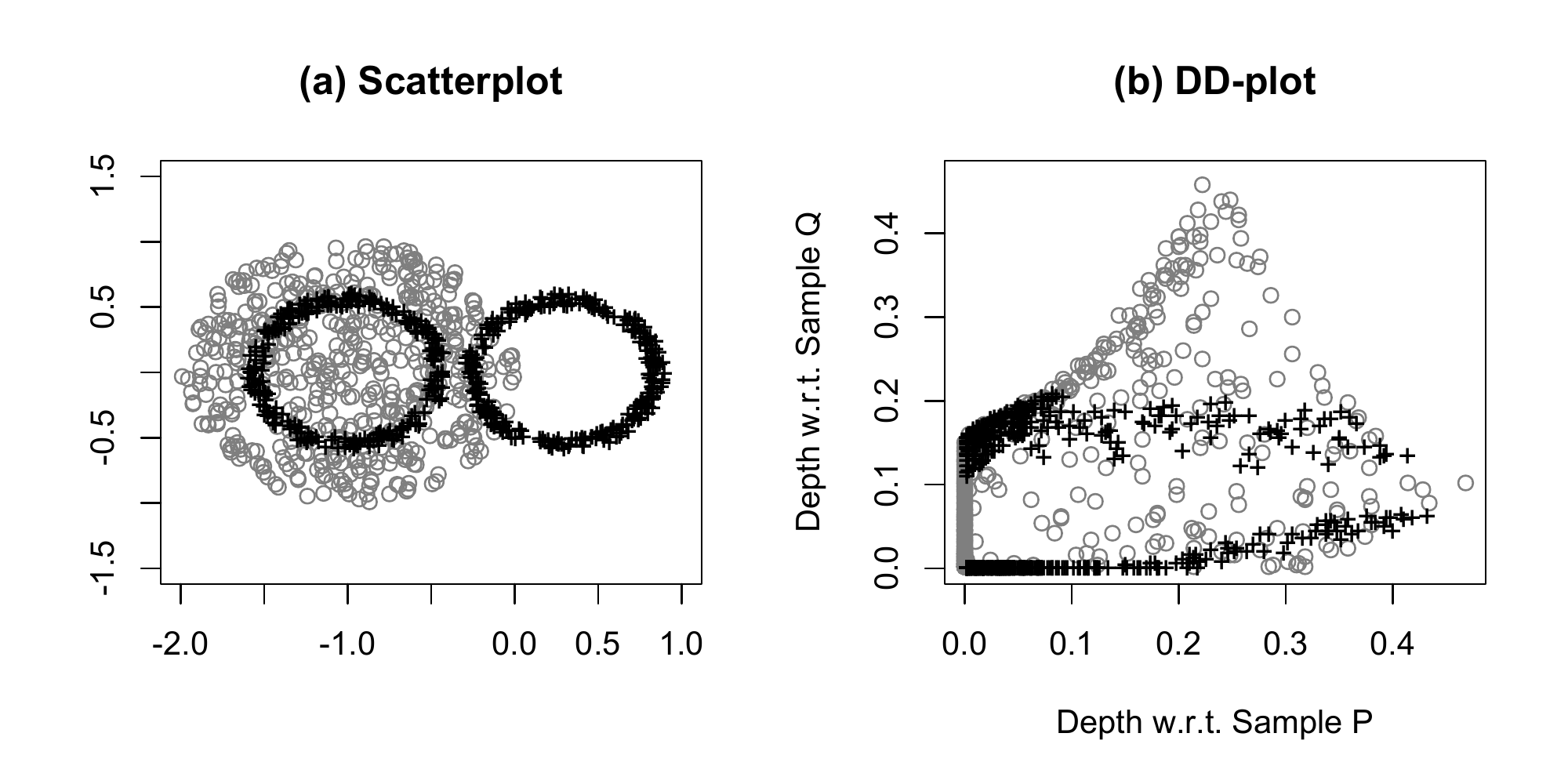} 

}

\end{knitrout}
\caption{Scatterplot of two uniform samples and associated DD--plot.}
\label{Fig2}
\end{figure}

The MD-classifier looks quite reasonable, but it has some drawbacks which are better understood with the help of the DD--plots. Those were introduced in  \cite{Liu1999} for graphical comparison of two multivariate distributions or samples  (see also \cite{Li2004}). 
Given two probability distributions,  \Prob \ and \ProbQ \ on $\Rea^p$, a DD--plot is a two-dimensional graph (regardless of $p$) in which, for every  $x \in \Rea^p$, the pair $(D_P(x),D_Q(x)) \in \Rea^2$ is represented. Examples of DD--plots appear in Figures \ref{Fig1} and \ref{Fig2}. Thus,  the MD-classifier gives to \ProbQ \ (resp. to \Prob) the points whose representation in the DD-plot is above (below) the main diagonal. 
Figure \ref{Fig1} contains two DD--plots corresponding to samples from bidimensional normal distributions, where \Prob, in both cases, is standard. The mean of \ProbQ \ in the first DD--plot is $(2,2)^t$ and its covariance is the identity. In the other case \ProbQ \ is centered but its covariance is twice the identity. In both graphs, points in black come from \Prob \ and points in gray from \ProbQ . We have employed the Halfspace Depth (HS) (see \cite{Liu1999}). All sample sizes are 500.
In both graphs the main diagonal is also drawn.

The MD-classifier is optimal in the first case, but it is plainly wrong in the second one since it classifies almost all points as produced by  \ProbQ. The idea developed  in \cite{Li2012} is that the DD--plot  contains information enabling a good classifier to be obtained. For instance, in the second DD--plot in Figure \ref{Fig1} the proportion of  gray points is very high in an area close to the vertical axis. Then, \cite{Li2012} proposed replacing the main diagonal by a function whose graph splits the DD--plot into two zones with the lowest misclassification rate (in that paper only the polynomial case is fully developed). This is termed the \emph{DD--classifier}.

The DD--classifier is a big improvement over the MD-classifier and, in the problem cited above, according to \cite{Li2012}, the DD--classifier gives a classification very close to the optimal one. However, an important limitation of the DD--classifier is that it is unable to deal efficiently with more than two groups. The solution of this problem for $g$ groups in \cite{Li2012} was to apply a majority voting scheme increasing the computational complexity with the need of solving $\binom{g}{2}$ two--groups problems.

Moreover, there are some two groups cases in which a function can not split the points in the DD--plot correctly. Let us consider the situation presented in Figure \ref{Fig2}. The points in the scatterplot come from two samples, with 2,000 points each. The gray points were taken from a uniform distribution, \ProbQ, on the unit ball centered on $(-1,0)^t$. The black points are from distribution  \Prob \ which is uniform on the union of two rings: a ring centered at $(-1,0)^t$ with inner (resp.\ outer) radius of $0.5$ (resp.\ $0.6$), and a ring of the same size centered at $(0.3,0)^t$. 
The optimal classifier assigns points in both rings to \Prob \ and the rest to \ProbQ. 
The associated DD--plot is also shown in Figure \ref{Fig2}. It is obvious that no function can split the DD--plot in two zones giving the optimal classifier since this would require a function separating  the points in areas with black points from the rest of the DD--plot, which is impossible. 
This problem, in this particular case, could be fixed by interchanging the axis. But, it is possible to  imagine a situation in which this rotation is not enough.

There are also several depths valid in functional spaces (we present some of them in Section \ref{sec:funcdepth}). Those depths can also be applied in classification problems making use of the DD-classifier, but suffering from the same problems we mentioned in the multidimensional case. Moreover, another limitation of the DD-plot is its incapability to take into account information coming from different sources. This fault is more important in the functional setting where some transformations of the original curves (such as derivatives) could be used for classification purposes simultaneously with the original trajectories. 
 
In this paper we present the DD$^G$--classifier as a way to fix all the  mentioned shortcomings of the DD--classifier  in the functional setting, although the procedure can also be applied to multivariate data or to the cartesian product of functional spaces with multivariate ones. In fact, the DD$^G$--classifier allows to handle more than two groups and also allows to incorporate information coming from different sources. The price we pay for this is an increment in the dimension which goes from 2 in the DD-plot to the number of groups times the number of different sources of information to be handled. The DD$^G$--classifier can also handle simultaneously more than one depth (increasing again the dimension). The letter $G$ in the name of the procedure makes reference to this incremented dimension. Finally, it allows to use regular classifiers (like kNN, SVM,\ldots). Since it is not longer compulsory to use functions to separate groups, then, for instance, it is possible to identify ``islands" inside a DD$^G$--plot, avoiding the need to use rotations.

Concerning the combination of information, it is worth to mention that, on one hand, in Section \ref{sec:funcdepth}, we  include some extensions of well known depths that allow  to construct new depths taking into account pieces of information from several sources; and, on the other hand, that some of the diagnostic tools of the classification procedures employed inside of the DD$^G$--classifier can be used to assess the relevance of the available  information. In order to avoid a too long paper, we only show this idea in the second example in Section \ref{Sec:Illustration} where we conclude that the relevant information is contained in the second derivative of the curves.    

The paper is organized as follows: in Section \ref{Sec.DDh_classifier} we present the basic ideas behind the proposed classifier. Section \ref{sec:funcdepth} is devoted to present some functional depths and to analyze some modifications which could improve them.  In Section \ref{Sec:Illustration} we show two examples of several classifiers applied to DD--plots. Section \ref{sec:simulaciones} contains the results of some simulations as well as applications to some real datasets. The paper ends with a discussion of the proposed method.

\section{DD$^G$-Classifier} \label{Sec.DDh_classifier}
In \cite{Li2012}, the DD--plot is defined, in the case in which only two groups are involved, as a two-dimensional graph where the pairs $(D_1(x),D_2(x))$ 
are plotted. Here, $D_i(x)$ is the depth of the point $x$ respect to the data in the $i$-th group.  With this notation, the DD--plot is, to put it simply, a map between the (functional) space $\mcal{X}$ where the data are defined, and $\R^2$:
\begin{eqnarray*}\label{map1}
x &\conv &(D_{1}(x),D_{2}(x)) \in \R^2.
\end{eqnarray*}

The  DD--classifier tries to identify the two groups using the information provided by the DD--plot. Since we have transformed our data to be in $\R^2$,  the task of separating classes is made in a much simpler framework, assuming that the depths contain relevant information about how to separate the groups. 
Thus, the choice of a depth has now become a crucial step. In \cite{Li2012} the classification rule was a polynomial function (up to a selected order $k$), ensuring that the point $\lrp{0,0}^t$ belongs to it. This rule has three main drawbacks. First, the number of different polynomials of order $k$ that can serve as a classification rule is  $\binom{N}{k}$, where $N$ is the sample size. This is the number of possible ways to select $k$ points from $N$, and each of the selections has an associated order $k$ polynomial which interpolates between these $k$ points and $(0,0)^t$. Clearly, as $N$ increases, the complexity of the estimation process grows at the rate $N^k$. Second, the problem of classifying more than two groups was solved in \cite{Li2012} using majority voting that needs to repeat the procedure for every combination of the groups. This means that the optimization must be solved $\binom{g}{2}$ times, where $g$ is the number of groups. Also, to avoid that the classification rule depends on the pre-specified order of the groups, the optimization procedure must be repeated interchanging the axes of the DD--plot. So, the number of polynomial models that must be computed to create the classification rule is $2\binom{g}{2}\binom{N}{k}$ that can be extremely large.  Finally, polynomials always give  borders between groups which do not allow the construction of zones assigned to one group included in a zone assigned to the other, like the horizontal black band between the gray zones in the DD--plot in Figure \ref{Fig2}. 

The DD$^G$--classifier which we propose here tries to offer a unified solution to these drawbacks. Suppose that we have a process in the product space $\mcal{X}=\mcal{X}_1\times\cdots\times\mcal{X}_p$, multivariate (functional) data, where we have $g$ groups (classes or distributions)  to be separated using data depths. Let us begin by assuming that $p=1$. The DD$^G$--classifier begins by selecting a depth $D$ and computing the following map:
\begin{eqnarray*}\label{map2.0}
x  &\conv & \mat{d}=({D}_1(x),\ldots,{D}_g(x)) \in \R^g.
\end{eqnarray*}
We can now apply any available  classifier that works in a $g$--dimensional space to separate the $g$ groups. The same idea is applied in \cite{Lange2014}. The main differences between \citeauthor{Lange2014}'s and our proposal are that in the former only finite-dimensional data are considered, and this map is a preliminary step to constructing what is called the feature space. Then, the authors only use a special kind of linear classifier on this feature space which requires making pairwise comparisons, thus classifying points using a majority vote scheme.  \cite{Mosler2014} apply this classifier to functional data, but only after performing a dimension-reduction technique to the data.
    
The extension of the procedure to the case $p>1$ is simple: we only need to select an appropriate depth $D^j$ for each subspace $\mcal{X}_j$ and consider the map

\begin{eqnarray*}\label{map2.1}
{\cal X} =\mcal{X}_1\times\ldots\times\mcal{X}_p & \conv & \R^{G}
\\
x =\lrp{x_1,\ldots,x_p} &\conv & \mat{d}=(\vec{D}^1(x_1),\cdots,\vec{D}^p(x_p)),
\end{eqnarray*}
where $\vec{D}^{i}(x_i)$ is the $g$-dimensional vector giving the depths of the point $x_i \in \mcal{X}_i$ with respect to the groups $1,\ldots,g$ and $G=g\times p$. 
    
Our last consideration is  related to the selection of the depth. As we stated before, the chosen depth may influence the result. The solution in \cite{Li2012} was to select the right depth by cross-validation. In principle, an obvious solution could be to include all the depths at the same time and, from the diagnostics of the classification method, select which depths are useful. But, this approach produces an increase of the dimension of vector $\mat{d}$ up to  $G=g\sum_{i=1}^p l_i$, where $l_i \geq 1$ is the number of   depths used in the $i$-th component. Clearly, the advantage of this approach depends on how the classification method can handle the information provided by the depths. Instead of  that, we propose  to select the useful depths  trying to maintain the dimension $G$ low. This choice can be done using the distance correlation $\mcal{R}$, see \cite{Szekely2007}, which characterizes independence between vectors of arbitrary finite dimensions. Recently, in \cite{Szekely2013}, a bias-corrected version was proposed. Here, our recommendation is to compute the bias-corrected distance correlation between the multivariate vector of depths ($\mat{d}$) and the indicator of the classes $\left( Y=\lrp{\ind{x\in C_1},\ind{x\in C_2},\ldots,\ind{x\in C_g}}\right)$, and select the depth that maximizes the distance correlation  among the available ones. In subsequent steps, other depths can be added having a low distance correlation between the new depth and those selected in previous steps. Also, using the recent extension of the distance correlation to functional spaces provided by \cite{Lyons2013}, this tool could be useful for assessing how much of the relation between the functional data and the indicator of the groups can be collected. Indeed, the computation of this measure is quite easy because it only depends on the distances among data (see Definition 4 in \cite{Szekely2007}). Later, in Section \ref{Sec:Illustration}, we  provide an example of the application of these ideas.

\subsection{Data Depths for Functional Data}
\label{sec:funcdepth}
As mentioned earlier, the DD--classifier  is especially interesting in the functional context because it enables the dimension of the classification problem to be decreased  from infinite to $G$. In this section, several functional data depths that will be  used later with the DD$^G$--classifier will be reviewed. Some  extensions to cover multivariate functional data are also provided.

\subsubsection{Fraiman and Muniz Depth (FM)} \label{subs.FMD}
The FM depth (\cite{Fraiman2001}) was the first one to be proposed in a functional context. It is also known as integrated depth by its definition. Given a sample $x_1,\ldots,x_N$ of functions defined on the interval $[0,T]$, let $S_t=\{x_1(t),\ldots,x_N(t)\}$ be the values of those functions on a given  $t\in[0,T]$. Denote by $F_{N,t}$, the empirical distribution of the sample $S_t$ and by $D_i(t)$ an univariate depth of $x_i(t)$ in this sample (in the original paper, $D_i(t)=1 -\left|1/2-F_{N,t}\left(x_i(t)\right)\right|$). Then the FM depth for the $i$-th datum is:
\begin{equation}\label{Eq.FM}
FM_i =\int_{0}^{T}{D_i(t)dt}.
\end{equation}

An obvious generalization of the FM depth is to consider different univariate depths to be integrated, like, for instance, the {\it Half Space depth} (HS, which is defined in (\ref{Eq.Depth1})), the {\it Simplicial depth} (SD) or the {\it Mahalanobis depth} (MhD):
\begin{eqnarray*}
D_i^{SD}(t)&=&2F_{N,t}\left(x_i(t)\right)\left(1-F_{N,t}\left(x_i(t)^-\right)\right),
\\
D_i^{MhD}(t)&=& \lrc{1+\lrp{x_i(t)-\hat{\mu}(t)}^2/\hat{\sigma}^2(t)}^{-1},
\end{eqnarray*}
where $\hat{\mu}(t), \hat{\sigma}^2(t)$ are estimates of the mean and variance at point $t$.

The choice of a particular univariate depth modifies the behavior of the FM depth. For instance, the deepest curve may vary depending on this selection. 

An interesting scenario arises when we are faced with multivariate functional data; i.e., when the elements belong to a product space of functional spaces:  $\mcal{X}=\mcal{X}^1\times\cdots\times\mcal{X}^p$. A depth combining the information of all components seems an appealing idea because it will maintain the dimension of our classification problem low, but it does so at the risk of losing some information. This can be done in the following two ways:

\begin{itemize}
\item \emph{Weighted depth}: given $x_i=(x_i^1,\ldots, x_i^p) \in \mcal{X}$, 
compute the depth of every component, obtaining the values ${FM} (x_i^j) , j=1,\ldots, p$, and then define a weighted version of the FM--depth (FM$^w$) as:
\[FM_i^w=\sum_{j=1}^p w_j {FM} (x_i^j),\] where $\vec{w}=\lrp{w_1,\ldots,w_p}$ is a suitable vector of weights. 
In the choice of $\vec{w}$, 
the  differences in the scales of the depths must be taken into account  (for instance, the FM depth using SD 
as the univariate depth takes values in $[0,1]$, whereas the Half Space depth always belongs to the interval $[0,1/2]$).

\item \emph{Common support}: suppose that all $\mcal{X}^i$ have the same support $\lrc{0,T}$ (this happens, for instance, when using the curves and their derivatives). In this case, we can define a $p$-summarized version of FM--depth (FM$^p$) depth as:
\[FM_i^p =\int_{0}^{T}D_i^p(t)dt,\] where $D_i^p(t)$ is a $p$-variate depth of the vector $\lrp{x_i^1(t),\ldots,x_i^p(t)}$ with respect to $S_t$. 

\end{itemize}

\subsubsection{{$h$}--Mode Depth ({$h$}M)}
The hM depth was proposed in \cite{Cuevas2007} as a functional generalization of the likelihood depth to measure how surrounded one curve is with respect to the others. The population hM depth of a datum $x_0$ is given by:
\[
f_h(x_0)=\E{K(m\lrp{x_0,X}/h)},
\]
where $X$ is a random element describing the population, $m$ is a suitable metric or semi-metric, $K(\cdot)$ is a kernel  and $h$ is the bandwidth  parameter.
Given a random sample $x_1,\ldots,x_N$ of $X$, the empirical \emph{h}--mode depth is defined as: 	
\begin{equation}\label{Eq.mode2}
\hat{f}_h(x_0)=N^{-1}\sum_{i=1}^{N}{K(m\lrp{x_0,x_i}/h)}.
\end{equation}
Equation \eqref{Eq.mode2} is similar to the usual nonparametric kernel density estimator,  with a main difference: as our interest is focused on what happens in a neighbourhood of each point, the bandwidth is not intended to converge to zero when $N \rightarrow \infty$, and the only constraint is that the bandwidth should be large enough to avoid pathological situations. For instance, the bandwidth should not be so small that every point in the sample has the same depth equal to $K(0)/N$. Our default choice for $h$ is the quantile 15\% of the distances among different points in the sample using as $K$ the standard gaussian density.  

A weighted depth of the components can  be  applied to use this depth with multivariate functional data. Another possibility in this case is to construct a new metric combining those defined in the components of the product space using a $p$-dimensional metric like, for example, the Euclidean; i.e., take 
\begin{equation}\label{Eq.L2Norm}
m\lrp{\lrp{x_0^1,\ldots,x_0^p},\lrp{x_i^1,\ldots,x_i^p}}:=\sqrt{m_1\lrp{x_0^1,x_i^1}^2+\cdots+m_p\lrp{x_0^p,x_i^p}^2},
\end{equation}
where $m_{i}$  denotes the metric in the $i$-component of the product space. It is important here to ensure that the different metrics of the spaces have  similar scales to avoid that one single component dominates the overall distance.    

\subsubsection{Random Projection Methods}
There are several depths based on random projections using basically the same scheme. 
Given a sample $x_1,\ldots,x_N$ of functions in a Hilbert space with scalar product $\inprod{\cdot }{\cdot }$, a unit vector $a$ in this space is randomly selected (independently of the $x_i$'s) and the data are projected onto the one-dimensional subspace generated by $a$. The sample depth of a datum $x$ is  the univariate depth of the  projection $\inprod{a}{x}$ with respect to the projected sample $\lrb{\inprod{a}{x_i}}_{i=1}^N$. Although theoretically a single projection is enough (see \cite{Cuesta-Albertos2007}), random projection methods usually generate several directions, $a_1, \ldots, a_R$, $R>1$ and summarizes them in different ways. Here, we will use:
\begin{itemize}
\item {\it Random Projection}
(RP): Proposed in \cite{Cuevas2007}, it uses univariate HS depth and summarizes the depths of the projections through the mean (using $R=50$ as a default choice). So, if $D_{a_r}(x)$ is the depth associated with the $r$--th projection, then \[RP(x)=R^{-1}\sum_{r=1}^R D_{a_r}(x).\] 
\end{itemize}   

The extensions to multivariate functional data are similar to those proposed for the FM depth, excepting for the fact that here, to use a $p$-variate depth with the projections, it is not required  all components to have a common support. The RPD depth proposed in \cite{Cuevas2007} is an example of this extension using the original curves and their derivatives as  components of  multivariate functional data, which in this case are two-dimensional.  

\subsubsection{Other Depth Measures}
Some other functional depth measures have been proposed in the last years although they are closely related with the three ones mentioned above. For instance, the \emph{Modified Band Depth} (MBD) proposed in \cite{Lopez-Pintado2009} can be seen as a particular case of the FM depth using as univariate depth, the simplicial one. The works by \cite{Ieva2013} and \cite{Claeskens2014} are in the same spirit as the extension of FM depth to multivariate functional data with common support. The first paper provides  a generalization of the MBD  that uses the Simplicial Depth as $p$-variate depth, and the second uses the multidimensional Half Space  depth.

The two proposals in \cite{Sguera2014} are the extension to  functional data of the multivariate spatial depth (see, e.g. \cite{Serfling2004}). The two depths, called Functional Spatial Depth (FSD) and Kernelized Functional Spatial Depth (KFSD), have different meanings. The first one is a global depth whereas the KFSD has a clear local pattern. We have tried them and we have obtained that FSD give results very similar to  FM or RP, while KFSD behaves as the hM depth. Because of this, we have included none of them in the simulations and real case studies.

\subsection{Classification Methods}
The last step in the DD$^G$--classifier procedure is to select a suitable classification rule.  Fortunately, we now have a purely multivariate classification problem in dimension $G$ and many procedures are known to handle it successfully based either in discriminant  or in regression ideas (see, for example, \cite{Ripley1996}).

Attending to their simplicity and/or easiness to draw inferences, we have selected  the following multivariate classification procedures to be used here:

\begin{enumerate}

\item \emph{Based on Discriminant Analysis}: The Linear Discriminant Analysis (LDA) is the most classical discriminant procedure. Introduced by Fisher, it is a particular application of the Bayes' Rule Classifier under the assumption that all the groups in the population have a normal distribution with different means,  but the same covariance matrix. The Quadratic Discriminant Analysis (QDA) is an extension relaxing the assumption of the equality among covariance matrices.

\item \emph{Based on Logistic Regression Models}: Here, the classifiers employ the logistic transformation to compute the posterior probability of belonging to a certain group using the information of the covariates. The Generalized Linear Models (GLM) combine linearly the information of vector $\mat{d}$, whereas the Generalized Additive Models (GAM) (see \cite{Wood2004}) relax the linearity assumption in GLMs allowing the use of a sum of general smooth functions of each variate. 

\item \emph{Nonparametric classification methods} are based on non-parametric estimates of the  densities of the groups. The most simple (and classical) one is the so-called $k$--Nearest Neighbour ($k$NN) in which, given $k \in \Nat$, the point $\mat{d}$ is assigned to the majority class of the $k$ nearest data points in the training sample. Another possibility is to estimate the probability of belonging to each group through the Nadaraya--Watson estimator using a common bandwidth for all data. This method will be denoted by NP. A $k$NN method could be considered an NP method using the uniform kernel and a locally selected bandwidth. These two methods are quite flexible and powerful but, unlike the previous ones, it is not easy to diagnose which part of the vector $\mat{d}$ is important for the final result.

\end{enumerate} 

There are many other classifiers that could be employed here, for instance: classification trees, artificial neural networks (ANN), support vector machines (SVM) or multivariate adaptive regression splines, \ldots but the application of any of these methods usually involves the choice of several auxiliary parameters or designs that must be tailored for every particular application. Also, as in the case of nonparametric classification methods, the trade--off between interpretability and predictability of these methods is biased to the latter. 

The choice among the different classifiers could be influenced by their theoretical properties and/or how easy it is to draw inferences. For example, from the theoretical point of view,  the $k$NN classifier can achieve optimal rates close to Bayes' risk (a complete review on this classifier can be found in \cite{Hall2008}) and it could be considered as the standard rule. But better inferences can be drawn from other classifiers such as LDA, GLM or GAM models.

\section{Illustration of Regular Classification Methods in DD--Plots}
\label{Sec:Illustration}

\subsection{Multivariate Example}
This section is devoted to explore the different classifiers that can be applied to DD--plots as an alternative to the proposal in \cite{Li2012}. In that paper, given $k_0 =1,2,\ldots$, the classifier is the polynomial $f$, with degree at most $k_0$ such that $f(0)=0$,  that gives the lowest misclassification error in the training sample. We denote this classifier by DD$k_0$. The candidate polynomials are constructed by selecting $k_0$ points of the sample and taking the polynomial going through these points and the origin.
In our implementation, we have ignored the step of selecting the order $k_0$ by cross--validation providing the best result for $k_0=1,2,3$ using, in each case, $M$ initial combinations  ($M=10,000$ by default) and optimizing the best $m$ ones ($m=1$ by default) following the implementation of \cite{Li2012}. Notice that the MD-classifier can be considered as a particular case of DD1, fixing the slope with a value of~1. 

\begin{figure}[htbp]
\centering
\begin{knitrout}\footnotesize
\definecolor{shadecolor}{rgb}{0.969, 0.969, 0.969}\color{fgcolor}

{\centering \includegraphics[width=.9\linewidth]{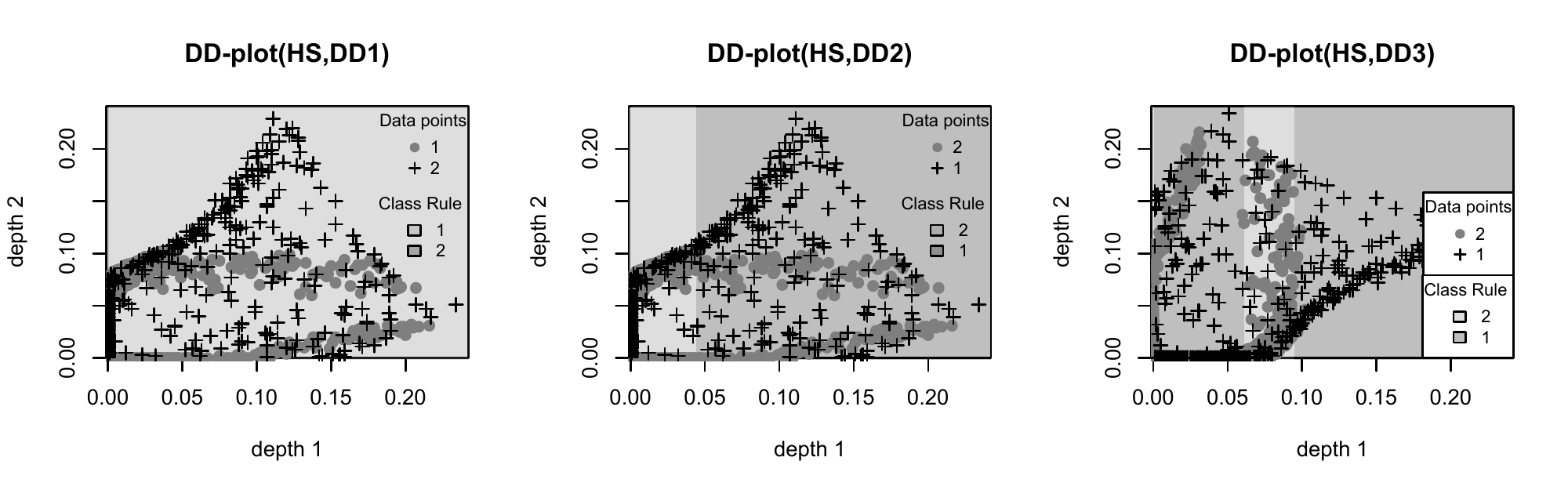} 

}

\end{knitrout}

\caption{From left to right DD--plot using DD1, DD2 and DD3 classifiers to the DD--plot in  Figure \ref{Fig2}(b). The depth in all cases is the HS.}\label{Fig3}
\end{figure}

The application to the example in Figure \ref{Fig2}(b) is plotted in Figure \ref{Fig3}, which shows the results for DD1, DD2 and DD3 classifiers. The titles of the subplots are in the general form DD--plot(\emph{depth}, \emph{classif}) where \emph{depth} is the depth employed (HS denotes the multidimensional Half Space depth) and \emph{classif} denotes the classification method. The sample points are colored gray or black to indicate the group they belong to. The background image is colored light gray and dark gray to indicate the areas where a new data point would be assigned to gray and black groups respectively.

The misclassification error rates are, respectively, (0.262, 0.215, 0.201). There is a clear superiority of  DD3  over the other classifiers but there are some areas (see for example, the rectangle $[0.0,0.2]\times [0.0,0.1]$) where a polynomial cannot satisfactorily classify the data. 

\begin{figure}[hbt]
\centering
\begin{knitrout}\footnotesize
\definecolor{shadecolor}{rgb}{0.969, 0.969, 0.969}\color{fgcolor}

{\centering \includegraphics[width=.9\linewidth]{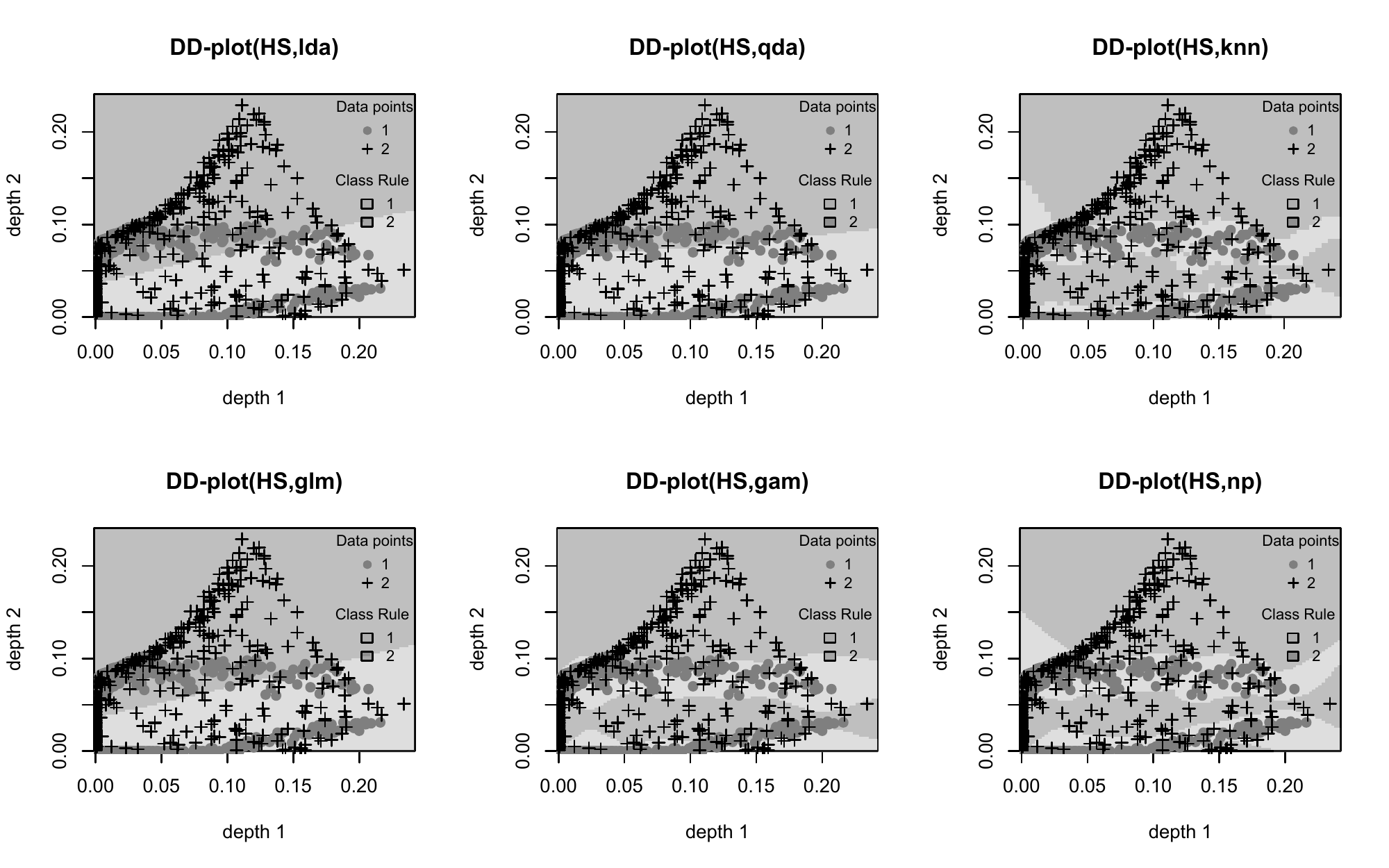} 

}

\end{knitrout}
\caption{From left to right, top to bottom DD--plot using LDA, QDA, $k$NN, GLM, GAM and NP classifiers to the DD--plot in  Figure \ref{Fig2}(b) The depth in all cases is the HS.}
\label{Fig4}
\end{figure}

Figure \ref{Fig4} shows the result to apply LDA, QDA, $k$NN, GLM,  GAM and NP to the same data. 
The misclassification  rates are, respectively, (0.472, 0.51, 0.136, 0.472, 0.152, 0.152). LDA, QDA and GLM methods do not achieve the result obtained by DD3 which is outperformed by $k$NN, GAM and NP. Notice that the optimal classifier gives a theoretical misclassification rate of $0.138$, very close to the result obtained with  $k$NN. The key of this improvement over the DD--classifier is the flexibility of $k$NN and GAM that can model complicated situations like this one.

\subsection{Functional Example: Tecator}
\begin{figure}[hbt]
\centering
\begin{knitrout}\footnotesize
\definecolor{shadecolor}{rgb}{0.969, 0.969, 0.969}\color{fgcolor}

{\centering \includegraphics[width=.9\linewidth]{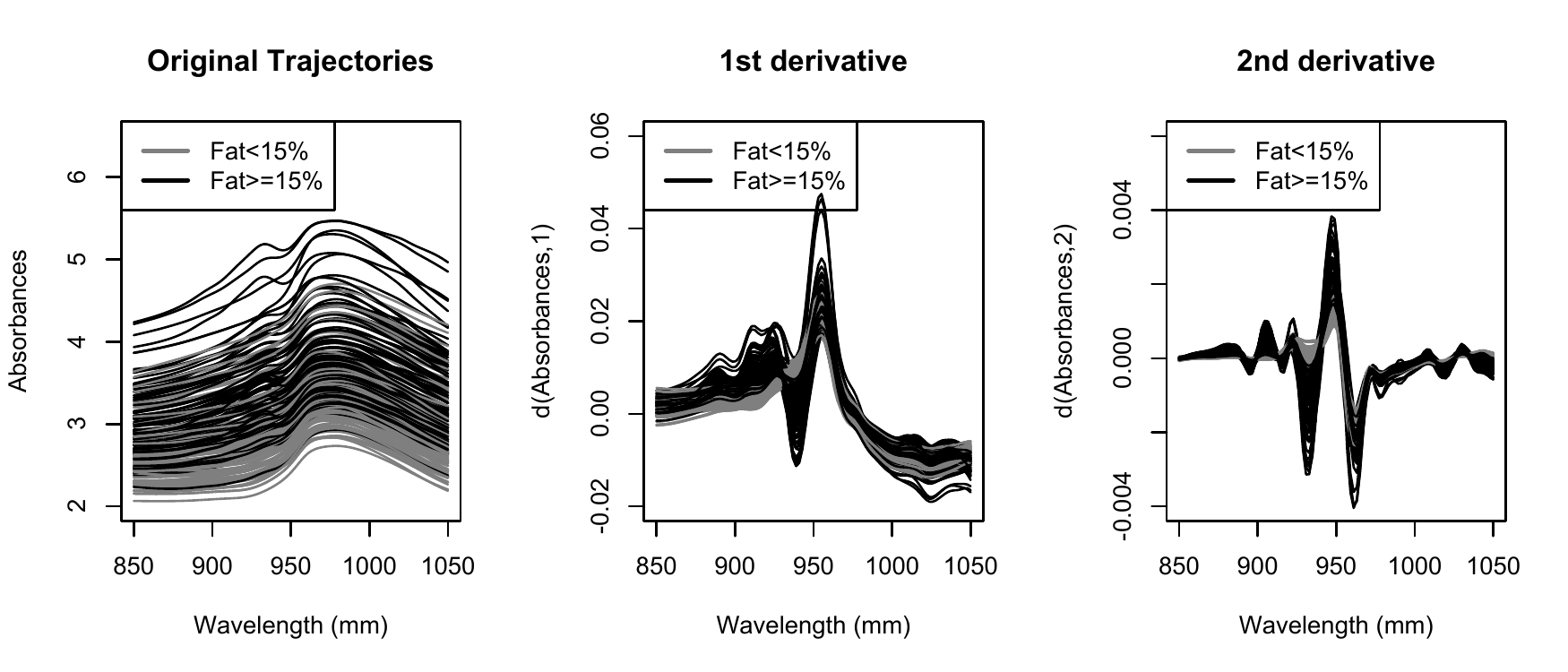} 

}

\end{knitrout}
\caption{Spectrometric curves of the Tecator dataset and their first two derivatives.}
\label{tecator}
\end{figure}

\begin{figure}[hbt]

\begin{knitrout}\footnotesize
\definecolor{shadecolor}{rgb}{0.969, 0.969, 0.969}\color{fgcolor}

{\centering \includegraphics[width=.9\linewidth]{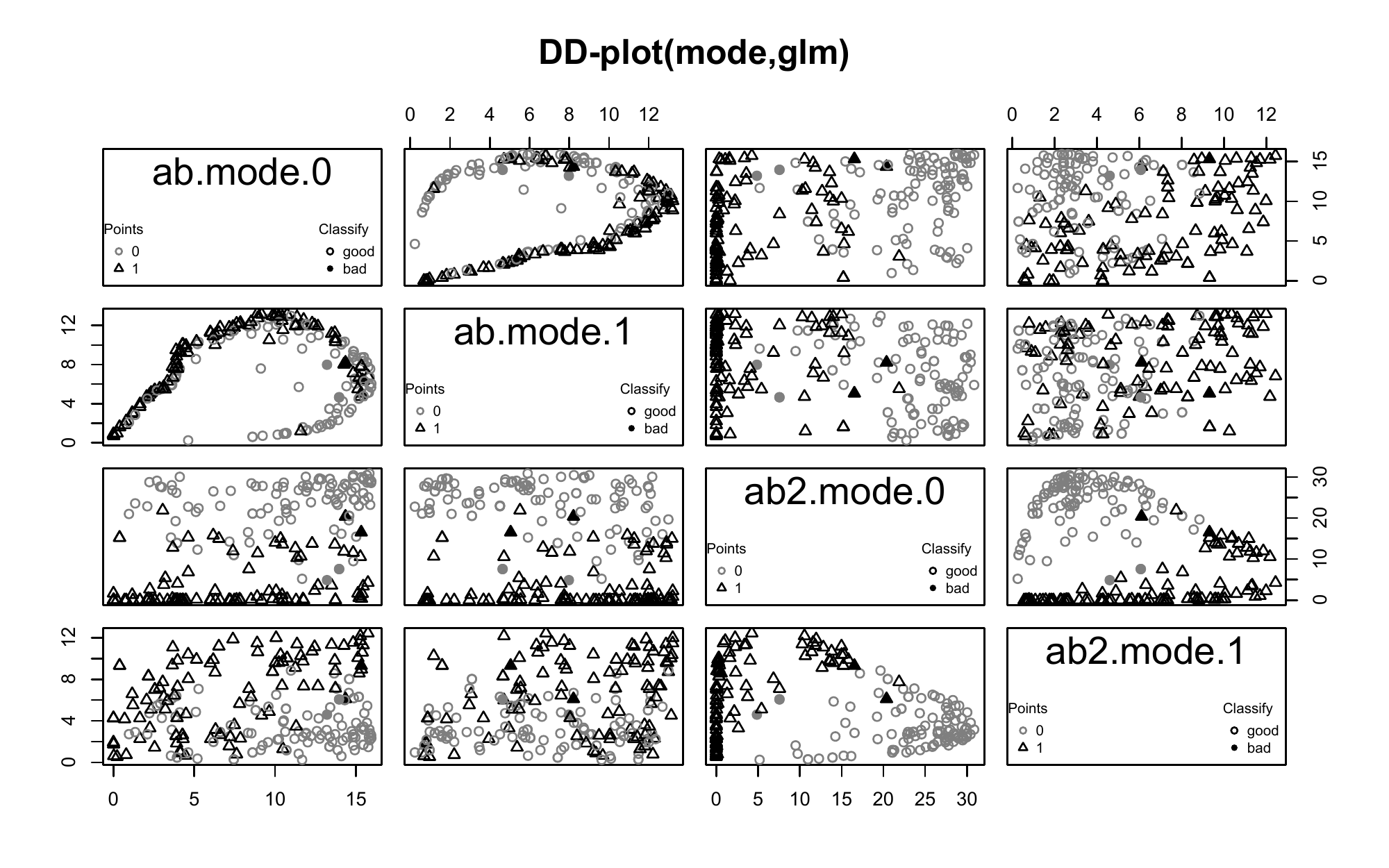} 

}

\end{knitrout}
\caption{Example of pairs of hM depths used by GLM classifier with the spectrometric curves of the Tecator dataset and its second derivative.}
\label{Fig5}
\end{figure}

In this section, we use the Tecator dataset to illustrate our procedure. 
Later, in Section \ref{subs.appl} this dataset will be revisited to compare the performance of the DD$^G$--classifier from the prediction point of view with other proposals.

Those data, which here are treated as a multivariate functional, were drawn for a spectrometric study where the goal was to predict the fat content of meat slices using absorbance curves provided by the device {\it Tecator Infrared Food Analyzer}. Many papers have treated those data  from the regression or the classification point of view (e.g., \cite{Ferraty2009}, \cite{Febrero2013} and references therein) with the conclusion that the relevant information for those goals is located  in the second derivative. Here, let us suppose that we are interested in identifying those samples with percentage of fat above 15\% (ifat=$\ind{\textrm{Fat}\ge 0.15}$) using the absorbance curves ($ab$) and their second derivatives ($ab2$) with the DD$^G$--classifier. First, concerning the depth, we use  {FM}, {RP} and {hM}, where we have employed the univariate Mahalanobis depth to compute the first two and the usual $L_2$-distance between functions in {hM} with the default choice of $h$ equal to the quantile 0.15 of the set $\lrb{d(x_i,x_j), i\ne j}$. Then, for each depth, at least five possibilities, identified through the different suffixes, can be explored:
\begin{itemize}
\item[\emph{.0}:] The depth uses only  original trajectories, $\mat{d}=\lrp{D_0^{0}(x),D_1^{0}(x)}$.
\item[\emph{.2}:] The depth only uses the second derivatives, $\mat{d}=\lrp{D_0^{2}(x),D_1^{2}(x)}$.
\item[\emph{.w}:] Use a weighted sum of the depth of the original trajectories and the depth of the second derivatives, $\mat{d}=\lrp{D_0^{w}(x),D_1^{w}(x)}$ with $D_i^w=0.5D_i^0+0.5D_i^2$.
\item[\emph{.m}:] Use all combinations depth/group,  $\mat{d}=\lrp{D_0^{0}(x), D_1^{0}(x), D_0^{2}(x), D_1^{2}(x)}$. 
\item[\emph{.p}:] The depths of the trajectories and their derivatives (a two dimensional functional dataset) are combined within the depth procedure. With FM and RP depths we use the two-dimensional Mahalanobis depth. The hM method uses an Euclidean metric as in (\ref{Eq.L2Norm}), $\mat{d}=\lrp{D_1^{p}(x),D_2^{p}(x)}$.  

\end{itemize}

\begin{table}[h]
\centering
\begin{tabular}{|c|r|r|r|r|r|}
\hline
 & FM.0 & FM.2 & FM.w & FM.m & FM.p \\\hline
$\mcal{R}(\textrm{ifat},\mat{d})$ & 0.058 & 0.771 & 0.393 & 0.365 & 0.058 \\\hline\hline
 & RP.0 & RP.2 & RP.w & RP.m & RP.p \\\hline
$\mcal{R}(\textrm{ifat},\mat{d})$ & 0.065 & 0.774 & 0.407 & 0.396 & 0.065 \\\hline\hline
 & hM.0 & hM.2 & hM.w & hM.m & hM.p \\\hline
$\mcal{R}(\textrm{ifat},\mat{d})$ & 0.114 & 0.789 & 0.706 & 0.762 & 0.114 \\\hline

\end{tabular}
\caption{Distance correlation between ifat and the different options for depths for the Tecator dataset.}
\label{dcor.tecator}
\end{table}

As mentioned above, the distance correlation proposed in \cite{Szekely2007} can help to detect the depth that best summarizes the  variate ifat. The distance correlation between the group variate (ifat) and the different depths are shown in Table \ref{dcor.tecator}. Since this metric only uses the distance among data, it can also be computed with respect to the functional covariates: $\mcal{R}(\mbox{ifat},ab)$=0.14, $\mcal{R}(\mbox{ifat},ab2)$=0.77,  supporting the idea that the important information for classification is contained in the second derivative. In Table \ref{dcor.tecator} we also see  that the depths based on the second derivative explain at least the same amount of information as the functional covariate does. In particular, FM.2, RP.2, hM.2, hM.w, hM.m have values over $0.7$. The first derivative ($ab1$) was not considered here because its distance correlation with ifat ($\mcal{R}(\mbox{ifat},ab1)$=0.63) is lower than the second one but both are quite related among them ($\mcal{R}(ab1,ab2)$=0.86). So, if we must select just one depth, the hM.2 must be the chosen one. In a second step, if we want to add more information, it is preferable to include the original trajectories because its lower distance correlation with $ab2$ ($\mcal{R}(ab,ab2)$=0.23).

The next step is to select a classifier that takes advantage of the dependence found by the distance correlation measure. The $k$NN could seem to be a good choice because it is quite simple to implement. But from the diagnosis point of view, a classifier like the GLM may be preferable. Using the hM.m depth (second best choice), we have four variates: $ab.mode.0$, $ab.mode.1$, $ab2.mode.0$, $ab2.mode.1$ where the notation $var.depth.group$ stands for the \emph{depth} computed for variate \emph{var} with respect to the points  in the group \emph{group}.

The result using a GLM  classifier is shown in Figure \ref{Fig5} with the combinations of the four variates, showing clearly that those associated with the second derivative  separate the two groups more efficiently. More interesting is that the contribution of each component can be assessed through the diagnosis of the GLM. The classical diagnosis of the estimates of a GLM model is shown in  Table \ref{Ex.tecator} where the variates associated with the depths of the second derivative are both clearly significant while this is not true for the original curves.

\begin{table}[h]
\centering
\begin{tabular}{|r|rrrl|}
\hline
& Estimate & Std. Error & $z$ value & $\PR(>|z|)$  \\\hline
(Intercept) & 3.538 & 2.161 & 1.637 & 0.102 \\
ab.mode.0 & \ensuremath{-0.473} & 0.166 & \ensuremath{-2.841} & 0.004 \\
ab.mode.1 & 0.054 & 0.155 & 0.347 & 0.729 \\
ab2.mode.0 & \ensuremath{-0.471} & 0.103 & \ensuremath{-4.585} & 0 \\
ab2.mode.1 & 1.09 & 0.301 & 3.624 & 0 \\ \hline
\end{tabular}
\caption{Output for the GLM classifier in the Tecator dataset.}
\label{Ex.tecator}
\end{table}

\section{A Simulation Study and the Analysis of Some Real Datasets} \label{sec:simulaciones} 
Four models (inspired by those in \cite{Cuevas2007}) were simulated in order to check the performance of the proposed classifier. In all cases, the curves are obtained from the process $X_{\cdot j}(t)=m_{j}(t)+e_{\cdot j}(t)$, where $m_{j}$ is the mean function of group $j=1,2$ and $e_{\cdot j}$ is a Gaussian process with zero mean and $Cov(e_{\cdot j}(s),e_{\cdot j}(t))=\theta_j \exp(-\left|s-t\right|/ 0.3)$. In all the models,  $\theta_1=0.5$ and $\theta_2=0.25$, giving the second group half the error of the first. The mean functions  include an additional parameter $k$ which is fixed at $k=1.1$. Note that  \cite{Cuevas2007} takes $k=1.2$ which makes the classification task easier due to a bigger separation of the groups. The functions were generated in the interval $\lrc{0,1}$ using an equispaced grid of $51$ points. These models were chosen trying to preserve a high similarity between groups jointly in the original trajectories and in their derivatives.

\begin{itemize}
\item \emph{Model 1}: The population $P_1$ has mean $m_1=30(1-t)t^{k}$. The mean for $P_2$ is $m_2=30(1-t)^kt$.

\item \emph{Model 2}: The population $P_1$ is the same as in  Model 1 but $P_2$ is composed of two subgroups as a function of a binomial variate $I$ with $\prob{I=1}=0.5$. Here,  $m_{2,I=0}=25(1-t)^kt$ and $m_{2,I=1}=35(1-t)^kt$.

\item \emph{Model 3}: Both populations are composed of two subgroups, with means $m_{1,I=0}=22(1-t)t^k$ and $m_{1,I=1}=30(1-t)t^k$, in the first population and  $m_{2,I=0}=26(1-t)^kt$ and $m_{2,I=1}=34(1-t)^kt$ in the second one. 

\item \emph{Model 4}: This uses the same subgroups defined in Model 3 but considers each subgroup as a group itself. So, this is an example with four groups. 
\end{itemize}

Thus, Models 1 and 4 are unimodal, while Models 2 and 3 contain at least one multimodal group. In last two models, the hM depth (which is local) should do better than the other ones.

\begin{figure}
\centering
\scalebox{0.9}{
\begin{knitrout}\footnotesize
\definecolor{shadecolor}{rgb}{0.969, 0.969, 0.969}\color{fgcolor}

{\centering \includegraphics[width=.9\textwidth]{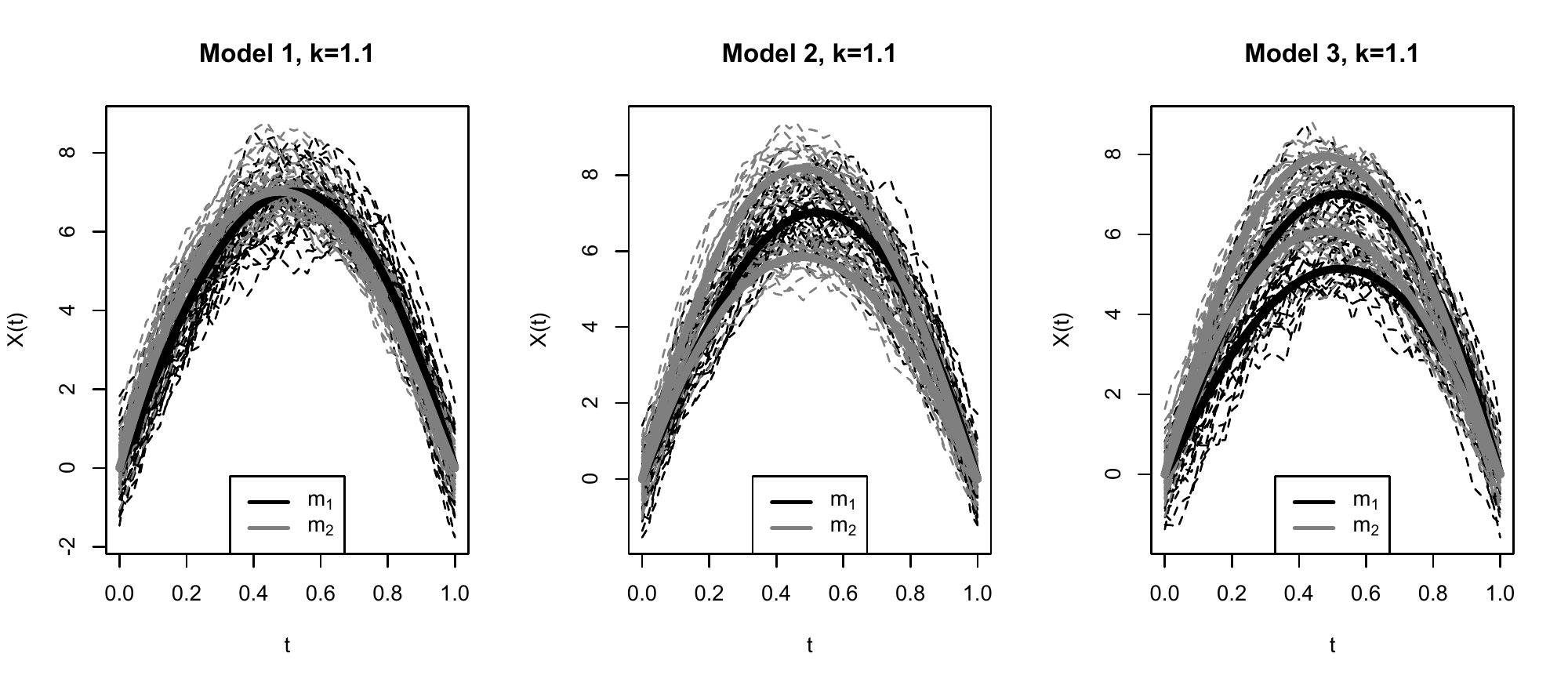} 

}

\end{knitrout}
}
\caption{A sample of $20$ functions for every simulation model along with the means of each sub-group ($m_1$'s (black lines) and $m_2$'s (gray lines)).}
\label{Fig.simus}
\end{figure}

The simulation results are based on 200 independent runs. In every run, $N=200$ training observations for Models 1 and 2 ($100$ for each group), and a test sample of $50$ observations from each group were generated. For Models 3 and 4, $N=400$ training observations are generated ($100$ for each subgroup). Tables \ref{tab.model1} to \ref{tab.model4} show the misclassification  rates for the test samples. Some curves obtained with each model are presented in Figure \ref{Fig.simus}.

For the comparison, the FM, RP and hM depths (computed with the default choices explained in Section \ref{Sec.DDh_classifier}) were employed using the original trajectories and/or the derivatives of every curve, which were computed using splines. The different depth options are denoted as in Section \ref{Sec:Illustration} except that the first derivative (\emph{.1}) is used instead of the second one. 

The distance $\mcal{R}$ is computed to select the best option from among the different depths (first row of Tables \ref{tab.model1} to \ref{tab.model4}). The overall winner is hM.w (closely followed by hM.p and hM.m) suggesting that the combined information of the curves and the first derivatives is better than using only one of them. This is a quite difficult example for a classification task as can be deduced from the relative small distance correlations obtained. As a reference, we have computed the Functional $k$NN (F$k$NN)  for all examples. 

The list of classifiers includes DD1, DD2 and DD3 as classical classifiers and also LDA, QDA, $k$NN, NP, GLM and GAM. Note that the procedures DD$i$, $i=1,2,3$ can not be used with the \emph{.m} option.

\setlength{\tabcolsep}{0.3mm}
\begin{table}[ht]
\centering
\begingroup\scriptsize
\begin{tabular}{c|ccccc|ccccc|ccccc|}
  & FM.0 & FM.1 & FM.w & FM.p & FM.m & RP.0 & RP.1 & RP.w & RP.p & RP.m & hM.0 & hM.1 & hM.w & hM.p & hM.m \\ 
  \hline
R(Y,d) & 0.23 & 0.28 & 0.34 & 0.32 & 0.32 & 0.34 & 0.25 & 0.36 & 0.36 & 0.38 & 0.38 & 0.42 & 0.50 & 0.49 & 0.47 \\ 
   \hline
DD1 & 27.2 & 21.5 & 20.3 & 21.2 &  & 24.5 & 16.7 & 17.6 & 17.8 &  & 20.3 & 15.9 & 15.6 & 15.7 &  \\ 
  DD2 & 24.9 & 20.4 & 18.6 & 19.4 &  & 24.0 & 15.4 & 17.4 & 17.6 &  & 18.7 & 13.7 & 12.6 & 12.8 &  \\ 
  DD3 & 25.1 & 20.4 & 18.7 & 19.5 &  & 24.4 & 15.8 & 17.1 & 17.3 &  & 19.0 & 14.0 & 13.1 & 13.3 &  \\ 
   \hline
LDA & 24.2 & 20.3 & 18.4 & 19.4 & 17.9 & 24.6 & 15.6 & 17.5 & 17.5 & 15.5 & 18.4 & 13.1 & 12.2 & 12.3 & 11.7 \\ 
  QDA & 24.5 & 20.3 & 18.4 & 19.5 & 18.1 & 25.1 & 15.8 & 17.5 & 17.6 & 16.4 & 18.5 & 13.2 & 12.1 & 12.3 & 11.9 \\ 
  kNN & 28.2 & 23.0 & 20.8 & 21.9 & 20.7 & 27.4 & 18.0 & 19.3 & 19.3 & 17.7 & 20.4 & 15.1 & 13.6 & 13.9 & 13.3 \\ 
  NP & 28.9 & 24.0 & 21.6 & 22.3 & 18.7 & 29.0 & 18.7 & 20.2 & 20.2 & 16.2 & 21.5 & 15.8 & 14.6 & 14.7 & 12.5 \\ 
  GLM & 24.1 & 19.9 & 18.3 & 19.2 & 17.8 & 24.3 & 15.6 & 17.2 & 17.2 & 15.4 & 18.4 & 13.1 & 12.1 & 12.3 & 11.6 \\ 
  GAM & 24.2 & 20.0 & 18.0 & 19.0 & 17.8 & 23.8 & 15.2 & 16.7 & 16.9 & 15.2 & 18.2 & 13.1 & 12.1 & 12.2 & 11.7 \\ 
   \hline
\end{tabular}
\endgroup
\caption{Distance correlation and misclassification rates for Model 1. Mean of $200$ runs.} 
\label{tab.model1}
\end{table}

The complete results for Model 1 are summarized in Table \ref{tab.model1} where the results of the distance correlation are, broadly speaking, confirmed: the best results are obtained with hM.m, closely followed by hM.w and hM.p. In these columns, 
the linear classifiers (LDA, GLM) seem to work slightly better than the others. This means that the simplest linear models are able to perform the classification task successfully. The F$k$NN was computed in its three versions: \emph{.0}, \emph{.1} and \emph{.p}  where the latter uses the euclidean distance combining the first two. The results obtained were, respectively, 23.04\%, 18.93\%, 19.02\%. 

\begin{table}[ht]
\centering
\begingroup\scriptsize
\begin{tabular}{c|ccccc|ccccc|ccccc|}
  & FM.0 & FM.1 & FM.w & FM.p & FM.m & RP.0 & RP.1 & RP.w & RP.p & RP.m & hM.0 & hM.1 & hM.w & hM.p & hM.m \\ 
  \hline
R(Y,d) & 0.24 & 0.16 & 0.22 & 0.32 & 0.26 & 0.28 & 0.13 & 0.24 & 0.32 & 0.24 & 0.48 & 0.34 & 0.50 & 0.44 & 0.48 \\ 
   \hline
DD1 & 32.5 & 26.6 & 25.5 & 16.1 &  & 31.1 & 22.3 & 22.0 & 21.6 &  & 14.0 & 15.8 & 10.6 & 10.2 &  \\ 
  DD2 & 22.8 & 27.1 & 21.0 & 16.0 &  & 20.8 & 21.1 & 16.6 & 16.3 &  & 11.5 & 16.0 & 10.6 & 9.9 &  \\ 
  DD3 & 23.0 & 27.4 & 21.3 & 16.1 &  & 20.8 & 20.9 & 16.6 & 16.3 &  & 11.8 & 16.2 & 10.8 & 10.2 &  \\ 
   \hline
LDA & 22.0 & 26.4 & 20.3 & 16.1 & 17.9 & 21.1 & 20.3 & 16.4 & 17.0 & 15.5 & 12.3 & 15.1 & 10.0 & 9.7 & 10.1 \\ 
  QDA & 22.3 & 26.7 & 20.6 & 15.9 & 18.7 & 20.8 & 20.5 & 16.1 & 16.5 & 16.0 & 11.9 & 14.9 & 9.8 & 9.3 & 9.8 \\ 
  kNN & 25.8 & 30.7 & 24.0 & 18.2 & 21.1 & 22.0 & 23.6 & 17.9 & 18.1 & 17.5 & 12.7 & 17.3 & 11.5 & 10.8 & 11.5 \\ 
  NP & 26.7 & 31.5 & 25.0 & 18.9 & 18.8 & 22.9 & 24.4 & 18.9 & 19.1 & 16.5 & 13.3 & 18.1 & 12.3 & 11.6 & 10.7 \\ 
  GLM & 22.1 & 26.3 & 20.2 & 15.6 & 17.9 & 19.7 & 20.1 & 15.4 & 15.9 & 15.0 & 11.7 & 15.1 & 9.7 & 9.3 & 9.7 \\ 
  GAM & 22.4 & 26.4 & 20.5 & 15.5 & 18.1 & 19.5 & 20.1 & 15.2 & 15.7 & 15.2 & 11.0 & 15.3 & 9.9 & 9.3 & 9.6 \\ 
   \hline
\end{tabular}
\endgroup
\caption{Distance correlation and misclassification rates for Model 2. Mean of $200$ runs.} 
\label{tab.model2}
\end{table}

 Model 2 (Table \ref{tab.model2}) is a difficult scenario for methods based on RP and FM depths as can be deduced from the low values of the distance correlation. These methods work well when the groups are homogeneous rather than being constituted of subgroups as in this case. The least misclassification error is obtained by the combinations hM.p--QDA, hM.p--GLM, hM.p--GAM ($9.3$\%), although many classifiers based on hM.w, hM.p or hM.m have misclassification  rates under 10\%. The results for F$k$NN were 13.63\%, 13.31\%, 12.57\%.


\begin{table}[ht]
\centering
\begingroup\scriptsize
\begin{tabular}{c|ccccc|ccccc|ccccc|}
  & FM.0 & FM.1 & FM.w & FM.p & FM.m & RP.0 & RP.1 & RP.w & RP.p & RP.m & hM.0 & hM.1 & hM.w & hM.p & hM.m \\ 
  \hline
R(Y,d) & 0.08 & 0.24 & 0.18 & 0.22 & 0.16 & 0.16 & 0.27 & 0.23 & 0.30 & 0.32 & 0.32 & 0.38 & 0.41 & 0.38 & 0.40 \\ 
   \hline
DD1 & 30.9 & 27.9 & 29.9 & 28.4 &  & 31.2 & 27.9 & 29.7 & 29.5 &  & 27.4 & 23.3 & 24.8 & 25.5 &  \\ 
  DD2 & 29.5 & 24.5 & 26.1 & 22.0 &  & 29.8 & 24.9 & 27.2 & 27.0 &  & 19.4 & 17.9 & 16.3 & 16.9 &  \\ 
  DD3 & 25.5 & 24.3 & 21.8 & 21.4 &  & 28.3 & 23.2 & 23.4 & 23.3 &  & 19.4 & 18.1 & 16.5 & 17.0 &  \\ 
   \hline
LDA & 32.0 & 25.2 & 29.8 & 26.4 & 25.1 & 32.0 & 27.0 & 30.3 & 30.6 & 27.0 & 24.0 & 18.4 & 19.1 & 20.2 & 18.2 \\ 
  QDA & 28.4 & 23.9 & 24.9 & 22.3 & 21.5 & 30.3 & 24.7 & 26.6 & 26.6 & 23.4 & 21.9 & 17.7 & 17.7 & 18.6 & 16.9 \\ 
  kNN & 25.9 & 25.0 & 22.2 & 21.3 & 21.6 & 27.3 & 23.6 & 22.9 & 22.9 & 22.4 & 20.3 & 18.2 & 16.6 & 17.2 & 16.7 \\ 
  NP & 25.5 & 24.3 & 22.1 & 21.3 & 21.3 & 27.7 & 23.3 & 22.5 & 22.8 & 21.7 & 19.9 & 17.9 & 16.4 & 17.0 & 16.4 \\ 
  GLM & 32.0 & 25.1 & 29.8 & 26.4 & 25.2 & 32.3 & 27.2 & 30.4 & 30.6 & 27.3 & 23.8 & 18.3 & 18.7 & 20.0 & 18.0 \\ 
  GAM & 24.8 & 23.8 & 21.2 & 20.6 & 21.6 & 26.1 & 22.9 & 22.1 & 21.8 & 21.8 & 19.4 & 17.6 & 16.2 & 16.8 & 16.2 \\ 
   \hline
\end{tabular}
\endgroup
\caption{Distance correlation and misclassification rates for Model 3. Mean of $200$ runs.} 
\label{tab.model3}
\end{table}

Model 3 (Table \ref{tab.model3}) is even harder for RP and FM methods. In both cases, the use of the first derivative is better than  the use of the original curves or a weighted version of them. For these depths, the best misclassification errors are obtained using the combined information (FM.p--GAM (20.6\%) and RP.m--NP (21.7\%)). This is also true for the hM method but it consistently yields lower misclassification errors. The best combinations are hM.w--GAM, hM.m--GAM ($16.2$\%) that are  better than the results using F$k$NN: 23\%, 21.4\%, 21.21\%.

\begin{table}[ht]
\centering
\begingroup\scriptsize
\begin{tabular}{c|ccccc|ccccc|ccccc|}
  & FM.0 & FM.1 & FM.w & FM.p & FM.m & RP.0 & RP.1 & RP.w & RP.p & RP.m & hM.0 & hM.1 & hM.w & hM.p & hM.m \\ 
  \hline
R(Y,d) & 0.60 & 0.47 & 0.65 & 0.65 & 0.63 & 0.60 & 0.56 & 0.67 & 0.69 & 0.66 & 0.64 & 0.58 & 0.69 & 0.68 & 0.68 \\ 
   \hline
DD1 & 21.6 & 29.1 & 17.8 & 18.1 &  & 23.9 & 19.5 & 16.9 & 16.8 &  & 19.5 & 17.9 & 14.6 & 14.4 &  \\ 
  DD2 & 21.7 & 28.9 & 17.2 & 17.7 &  & 23.9 & 18.9 & 16.7 & 16.4 &  & 17.9 & 16.6 & 12.7 & 12.5 &  \\ 
  DD3 & 23.0 & 29.6 & 18.9 & 19.2 &  & 25.2 & 20.3 & 18.3 & 18.1 &  & 19.4 & 18.0 & 14.6 & 14.4 &  \\ 
   \hline
LDA & 21.0 & 27.4 & 16.4 & 16.9 & 16.8 & 23.2 & 18.3 & 15.9 & 16.0 & 14.3 & 17.6 & 15.9 & 12.5 & 12.1 & 12.0 \\ 
  QDA & 21.0 & 28.0 & 17.0 & 18.4 & 17.4 & 23.8 & 18.9 & 16.6 & 16.5 & 16.3 & 17.9 & 16.2 & 11.8 & 12.2 & 12.7 \\ 
  kNN & 21.5 & 30.2 & 17.1 & 17.6 & 17.4 & 23.0 & 19.2 & 16.1 & 16.0 & 15.3 & 17.3 & 16.5 & 12.0 & 12.2 & 12.4 \\ 
  NP & 20.7 & 28.2 & 16.6 & 17.1 & 17.0 & 22.3 & 18.6 & 15.8 & 15.6 & 15.3 & 17.0 & 16.1 & 11.9 & 12.0 & 12.4 \\ 
  GLM & 20.9 & 27.7 & 15.9 & 16.4 & 16.1 & 23.0 & 18.0 & 15.5 & 15.4 & 14.0 & 16.6 & 15.8 & 11.4 & 11.3 & 11.3 \\ 
  GAM & 20.6 & 27.5 & 16.0 & 16.5 & 16.9 & 21.8 & 17.9 & 15.1 & 15.1 & 14.5 & 15.9 & 15.8 & 11.3 & 11.5 & 12.1 \\ 
   \hline
\end{tabular}
\endgroup
\caption{Distance correlation and misclassification rates for Model 4. Mean of $200$ runs.} 
\label{tab.model4}
\end{table}

The results for Model 4 (Table \ref{tab.model4}) are better than those for Model 3, supporting the idea that homogeneous groups are easier to classify with RP and FM depths. In all cases, the weighted version improves the classification of each component alone. This hints that the two components have complementary pieces of the information needed for classification. The best combinations for each depth are: FM.w--GLM (15.9\%), RP.m--GLM (14\%) and  hM.w--GAM, hM.p--GLM, hM.m--GLM ($11.3$\%). The F$k$NN  gives quite disappointing results:  19.8\%, 21.45\%, 20.16\%, probably due to the difficulty of the scenario.

\subsection{Application to Real Datasets}\label{subs.appl}
We have  applied our proposal to several popular datasets in the functional data analysis literature. A nice review on functional classification can be seen in \cite{Baillo2010}. In that follows, we will briefly describe the datasets, the results found in the literature and our best results using DD$^G$--classifier.

\begin{itemize}
\item \emph{Tecator}: When the Tecator dataset is used for classification, several differences in the scheme employed can be found in the literature;  including the cutoff for groups, the size of the training and testing samples and even the number of runs. In \cite{Febrero2012}, the scheme cutoff=15\% /train=165/test=50/runs=500 is employed with a best result of a FKGAM model of 2.1\% of misclassification error. Here, using depths, the best result is $1.3$\% with the hM.2--DD2 model. The classical F$k$NN using the second derivative obtains $1.9$\%.

In \cite{Galeano2015} a misclassification error of 1\% is reported using a centroid method with the functional Mahalanobis semidistance and with the scheme cutoff=20\%/train=162 /test=53/runs=500. Following the same scheme but with $200$ runs,  the hM.2--DD2 (error rate: $1.3\%$) performs quite well and slightly better than the classifier using $k$NN ($2.5\%$). In fact, all the classifiers using the second derivative show misclassification rates in the interval [$1.3$\%, $3.3$\%]; that can be compared with the F$k$NN classifier that obtains $1.92$\%.

\item \emph{Berkeley Growth Study}: This dataset contains the heights of 39 boys and 54 girls from age 1 to 18. It constitutes a classical example included in \cite{Ramsay2005} and in the \texttt{fda} R-package. 

As a classification problem, this dataset was treated in \cite{Baillo2008}, where using a F$k$NN procedure, a best cross-validation missclassification rate of $3.23\%$ was obtained. In our application, the best result is obtained by the combinations hM.0--LDA, hM.0--QDA with $2.2\%$.

\item \emph{Phoneme}: The \emph{phoneme} dataset is also quite popular in the FDA community although its origins are in the area of Statistical Learning (see \cite{Hastie1995}). The dataset has 2000 log-periodograms of 32ms duration corresponding to five different phonemes (\emph{sh, dcl, iy, aa, ao}). 

It appeared as a functional classification problem in \cite{Ferraty2003}. Randomly splitting the data into training and test samples with $250$ cases, $50$ per class, in each sample,  and repeating the procedure $200$ times, the best result  achieved by the authors was an $8.5\%$ misclassification rate. With our proposals, the combination hM.m--LDA misclassifies  $7.5\%$. 

This dataset was also used in \cite{Delaigle2012} but it was restricted to the use of the first 50 discretization points and to the binary case using the two most difficult phonemes, (\emph{aa, ao}), obtaining a misclassification rate of $20\%$ when  $N=100$. Our best result is $18.6\%$ obtained by hM.w--QDA although most hM procedures yield errors below $20\%$.

\item \emph{MCO Data}: These curves correspond to mitochondrial calcium overload (MCO), measured every 10 seconds for an hour in isolated mouse cardiac cells. The data (two groups: control and treatment) were used as functional data in \cite{Cuevas2004} for ANOVA testing and the dataset is available in the \texttt{fda.usc} package. 

As an FDA classification problem, it was considered in \cite{Baillo2008} where using a cross validation procedure, a best error rate of $11.23\%$ was obtained. Our best results are the combinations hM.1--DD1, hM.m--LDA, hM.m--QDA, hM.m--NP with an error rate of $2.2\%$.

\item \emph{Cell Cycle}: This dataset contains temporal gene expression measured every 7 minutes (18 observations per curve) of $90$ genes involved in the yeast cell cycle. The data were originally obtained by \cite{Spellman1998} and used in \cite{Leng2006} and \cite{Rincon2012} with the goal of classifying these genes into two groups. The first group has $44$ elements related with G1 phase regulation. The remaining 46 genes make up the second group and are related to the S, S/G2, G2/M and M/G1 phases. The dataset has several missing observations which were imputed in this work using a B-spline basis of $21$ elements. 

Both papers cited above obtain a misclassification rate of $10\%$ (9 misclassified genes) but with different number of errors for each group. Our proposal achieves a $6.7\%$ rate with the combinations  hM.1--DD1, hM.w--kNN, hM.w--NP, hM.p--DD1, hM.m--kNN, hM.m--NP but almost all procedures based on hM.1 or hM.w yield a misclassification rate of $8.9\%$ at most.

\item \emph{Kalivas}: This example comes from \cite{Kalivas1997}. It was used for  classification  in \cite{Delaigle2012}. It contains near-infrared spectra of 100 wheat samples from 1100nm to 2500nm in 2nm intervals. Two groups are constructed using the protein content of each sample, using a binary threshold of $15\%$ that places $41$ data in the first group and $59$ in the second. 

Our best result for $200$ random samples of size $50$ was the combination FM.m--QDA with a  $3.7\%$ misclassification error. This rate is quite far from the best in \cite{Delaigle2012} ($\mathit{CENT}_{PC1}=0.22\%$) using the centroid classifier but the latter requires projecting in a specific direction that in this case corresponds to small variations on the subinterval $[1100,1500]$. Notice that any depth procedure based on the whole interval cannot achieve a better result than a technique focused in the small interval that contains the relevant information for the discrimination process. 

\end{itemize}

\section{Conclusions}
In this paper we  present a procedure that extends the DD-classifier proposed in \cite{Li2012} and adapts it to the functional context in several ways:
\begin{itemize}

\item Due to the flexibility of the new classifiers considered, the  proposal can deal with several depths or with more than two groups in the same 
integrated framework. In fact, the DD$^G$ classifier converts the data into a multivariate dataset  whose columns are constructed using depths and the new classifiers are classical multivariate classifiers based on discrimination  (LDA, QDA) or regression procedures ($k$NN, NP, GLM, GAM). More classifiers could be considered here (such as SVM or ANN) without changing so much the procedure. 
The choice of a classifier must be based on the weakness and the strengths of each one. For instance, for the diagnostic part it is recommended the use of  classifiers as LDA or GLM because it is easier to interpret the rule for separating groups perhaps with some cost on the predictive performance.

\item
The DD$^G$--classifier  is especially interesting in a high-dimensional or functional framework because it changes the dimension of the classification problem from large or infinite to $G$, where $G$ depends only on the number of groups under consideration and the number of depths that the statistician decides to employ, perhpaps, times the number of sources of information to be used. For instance, if we have 3 groups in the data and the method is using 2 different depths, the multivariate dimension of the DD$^G$--classifier is 6. Clearly, this is a more tractable dimension for the problem, but there are, in addition, some ways to reduce this number. In this paper, a review of functional data depths is made by including modifications to summarize multivariate functional data (the data are made up of vectorial functions) without increasing or even reducing the dimension of the problem at hand.

In a multivariate setting, this might not be so advantageous because the dimension $G$ is a multiple of the number of groups and it could sometimes be greater than the dimension of the original space. For instance, in the classical example of Fisher Iris data, there are four variables and three groups, so that using the DD$^G$--classifier map in its most simple case can be worked in dimension three. But  we can also consider a univariate depth for each variable and then the dimension $G$ grows up to twelve. 
\item The execution time for each method, measured in CPU seconds, depends on the complexity of the combination depth/classifier. Taking the Model 1 with the original curves as a reference, the fastest time is obtained by the combination FM--LDA (0.05s). Similar times are obtained by QDA and GLM methods. The use of the GAM classifier adds 0.07s. The nonparametric classifiers (NP, $k$NN) typically add 0.35--0.40s to the time due to the computation of the distance matrix among points in the DD--plot. The use of random projections increases the time 0.01s per combination and the computation of the hM depth takes 1.05s which is the time employed by the F$k$NN. The use of a combined depth option (\emph{.w}, \emph{.p}, \emph{.m}) doubles the execution time. The DD${k}$ choices obtain 0.07, 13.77 and 39.77s, respectively, with the default choice ($M=10000$, $m=1$) even though better execution times can be achieved using $M=500$, $m=50$ maintaining similar misclassification rates.    
\item The functions needed to perform this procedure are freely available at CRAN in the \texttt{fda.usc} package (\cite{Febrero2012}) in versions higher than 1.2.2. \texttt{classif.DD} is the principal function  and contains all the options shown in this paper related to depths and classifiers. Most figures we present are   regular outputs of this function.      
\end{itemize}

\begin{center}
{\bf SUPPLEMENTAL MATERIALS}
\end{center}
\begin{description}
\item[Supplemental Code:] Rar compressed file containing the code with the plots and results in the paper (paper.code.R), the code of the simulation studies (simul.xxx.R) and the code for applications to real datasets (classif.xxxx.R). A folder with the data is also included (rar file). 
\end{description}

\begin{center}
	{\bf SUPPLEMENTAL MATERIALS}
\end{center}
\begin{description}
	\item[Supplemental Code:] Rar compressed file containing the code with the plots and results in the paper (paper.code.R), the code of the simulation studies (simul.xxx.R) and the code for applications to real datasets (classif.xxxx.R). A folder with the data is also included (rar file). 
\end{description}


\end{document}